\input harvmac
\noblackbox
\newcount\figno
\figno=0
\def\fig#1#2#3{
\par\begingroup\parindent=0pt\leftskip=1cm\rightskip=1cm\parindent=0pt
\baselineskip=11pt \global\advance\figno by 1 \midinsert
\epsfxsize=#3 \centerline{\epsfbox{#2}} \vskip 12pt {\bf Fig.
\the\figno:} #1\par
\endinsert\endgroup\par
}
\def\figlabel#1{\xdef#1{\the\figno}}
\def\encadremath#1{\vbox{\hrule\hbox{\vrule\kern8pt\vbox{\kern8pt
\hbox{$\displaystyle #1$}\kern8pt} \kern8pt\vrule}\hrule}}

\input epsf

\overfullrule=0pt

%
\def\tilde{\widetilde}

%
\def\inbar{\,\vrule height1.5ex width.4pt depth0pt}
\def\IB{\relax{\rm I\kern-.18em B}}
\def\IC{\relax\hbox{$\inbar\kern-.3em{\rm C}$}}
\def\ID{\relax{\rm I\kern-.18em D}}
\def\IE{\relax{\rm I\kern-.18em E}}
\def\IF{\relax{\rm I\kern-.18em F}}
\def\IG{\relax\hbox{$\inbar\kern-.3em{\rm G}$}}
\def\IH{\relax{\rm I\kern-.18em H}}
\def\II{\relax{\rm I\kern-.18em I}}
\def\IK{\relax{\rm I\kern-.18em K}}
\def\IL{\relax{\rm I\kern-.18em L}}
\def\IM{\relax{\rm I\kern-.18em M}}
\def\IN{\relax{\rm I\kern-.18em N}}
\def\IO{\relax\hbox{$\inbar\kern-.3em{\rm O}$}}
\def\IP{\relax{\rm I\kern-.18em P}}
\def\IQ{\relax\hbox{$\inbar\kern-.3em{\rm Q}$}}
\def\IR{\relax{\rm I\kern-.18em R}}
\font\cmss=cmss10 \font\cmsss=cmss10 at 7pt
\def\IZ{\relax\ifmmode\mathchoice
{\hbox{\cmss Z\kern-.4em Z}}{\hbox{\cmss Z\kern-.4em Z}}
{\lower.9pt\hbox{\cmsss Z\kern-.4em Z}} {\lower1.2pt\hbox{\cmsss
Z\kern-.4em Z}}\else{\cmss Z\kern-.4em Z}\fi}
\def\IGa{\relax\hbox{${\rm I}\kern-.18em\Gamma$}}
\def\IPi{\relax\hbox{${\rm I}\kern-.18em\Pi$}}
\def\ITh{\relax\hbox{$\inbar\kern-.3em\Theta$}}
\def\IOm{\relax\hbox{$\inbar\kern-3.00pt\Omega$}}

\font\zfont = cmss10 
 
\def\bigone{\hbox{1\kern -.23em {\rm l}}}
\def\ZZ{\hbox{\zfont Z\kern-.4emZ}}

\def\G{\Gamma}
\def\a{\alpha}
\def\b{\beta}
\def\d{\delta}

\def\k{\kappa}
\def\l{\lambda}

\def\o{\over}

\def\IR{\relax{\rm I\kern-.18em R}}
\def\I1{\relax{\rm I\kern-.6em 1}}
\def\Dsl{\,\raise.15ex\hbox{/}\mkern-13.5mu D}
\def\Gsl{\,\raise.15ex\hbox{/}\mkern-13.5mu G}
\def\Csl{\,\raise.15ex\hbox{/}\mkern-13.5mu C}
\font\cmss=cmss10 \font\cmsss=cmss10 at 7pt
\def\pa{\partial}

\def\co{{\cal O}}

\font\zfont = cmss10 
 
\def\bigone{\hbox{1\kern -.23em {\rm l}}}
\def\ZZ{\hbox{\zfont Z\kern-.4emZ}}

\def\unlockat{\catcode`\@=11}
\def\lockat{\catcode`\@=12}

\unlockat

\def\newsec#1{\global\advance\secno by1\message{(\the\secno. #1)}
\global\subsecno=0\global\subsubsecno=0\eqnres@t\noindent
{\bf\the\secno. #1} \writetoca{{\secsym}
{#1}}\par\nobreak\medskip\nobreak}
\global\newcount\subsecno \global\subsecno=0
\def\subsec#1{\global\advance\subsecno  
by1\message{(\secsym\the\subsecno. #1)}
\ifnum\lastpenalty>9000\else\bigbreak\fi\global\subsubsecno=0
\noindent{\it\secsym\the\subsecno. #1} \writetoca{\string\quad
{\secsym\the\subsecno.} {#1}}
\par\nobreak\medskip\nobreak}
\global\newcount\subsubsecno \global\subsubsecno=0
\def\subsubsec#1{\global\advance\subsubsecno by1
\message{(\secsym\the\subsecno.\the\subsubsecno. #1)}
\ifnum\lastpenalty>9000\else\bigbreak\fi
\noindent\quad{\secsym\the\subsecno.\the\subsubsecno.}{#1}
\writetoca{\string\qquad{\secsym\the\subsecno.\the\subsubsecno.}{#1}}
\par\nobreak\medskip\nobreak}
\def\subsubseclab#1{\DefWarn#1\xdef
#1{\noexpand\hyperref{}{subsubsection}%
{\secsym\the\subsecno.\the\subsubsecno}%
{\secsym\the\subsecno.\the\subsubsecno}}%
\writedef{#1\leftbracket#1}\wrlabeL{#1=#1}}
\lockat

\def\IP{\relax{\rm I\kern-.18em P}}
\def\IC{\relax{\rm I \kern-.5em C}}
\def\IZ{\relax{\rm I\kern-.18em Z}}
\lref\cntc{
M.~Cederwall, B.E.W.~Nilsson and D.~Tsimpis,
``Spinorial cohomology and maximally supersymmetric theories,''
JHEP {\bf 0202}, 009 (2002)
[arXiv:hep-th/0110069].
}
%
%
\lref\martin{
M.~Cederwall,
``Superspace methods in string theory, supergravity and gauge theory,''
arXiv:hep-th/0105176.
}
%
\lref\cntb{
M.~Cederwall, B.E.W.~Nilsson and D.~Tsimpis,
``D = 10 superYang-Mills at ${\cal O}(\alpha^{\prime 2})$,''
JHEP {\bf 0107}, 042 (2001)
[arXiv:hep-th/0104236].
}
%
\lref\cnta{
M.~Cederwall, B.E.W.~Nilsson and D.~Tsimpis,
``The structure of maximally supersymmetric Yang-Mills theory:  Constraining higher-order corrections,''
JHEP {\bf 0106}, 034 (2001)
[arXiv:hep-th/0102009].
}
%
\lref\at{
O.~D.~Andreev and A.~A.~Tseytlin,
``Partition Function Representation For The Open Superstring Effective Action: Cancellation Of Mobius Infinities And Derivative Corrections To
Born-Infeld Lagrangian,''
Nucl.\ Phys.\ B {\bf 311}, 205 (1988).
}
%
\lref\ga{
S.~J.~Gates, K.~S.~Stelle and P.~C.~West,
``Algebraic Origins Of Superspace Constraints In Supergravity,''
Nucl.\ Phys.\ B {\bf 169}, 347 (1980).
}
%
\lref\gb{
S.~J.~Gates and W.~Siegel,
Nucl.\ Phys.\ B {\bf 163}, 519 (1980).
}
%
%
\lref\nilssona{
B.E.W.~Nilsson, 
``Off-shell fields for the
Ten-Dimensional Supersymmetric Yang-Mills Theory,'' 
G\"{o}teborg-ITP-81-6; 
``Pure Spinors As Auxiliary Fields 
In The Ten-Dimensional Supersymmetric Yang-Mills Theory,''
Class.\ Quant.\ Grav.\  {\bf 3}, L41 (1986).
}
%
%
\lref\fosse{
L.~De Fosse, P.~Koerber and A.~Sevrin,
``The uniqueness of the Abelian Born-Infeld action,''
Nucl.\ Phys.\ B {\bf 603}, 413 (2001)
[arXiv:hep-th/0103015].
}
%
\lref\ulf{
U.~Gran,
``GAMMA: A Mathematica package for performing 
Gamma-matrix algebra and  Fierz transformations in arbitrary dimensions,''
arXiv:hep-th/0105086.
}
%
\lref\tseytlin{
A.~A.~Tseytlin,
``Born-Infeld action, supersymmetry and string theory,''
arXiv:hep-th/9908105.
}
\lref\lie{
A.~M.~Cohen, M.~van Leeuwen and B.~Lisser,
 LiE v. 2.2 (1998),
http://wallis.univ-poitiers.fr/maavl/LiE/
}
%
%
\lref\bi{
M.~Born and L.~Infeld,
``Foundations Of The New Field Theory,''
Proc.\ Roy.\ Soc.\ Lond.\ A {\bf 144}, 425 (1934).
}
%
%
\lref\fta{
E.~S.~Fradkin and A.~A.~Tseytlin,
``Quantum String Theory Effective Action,''
Nucl.\ Phys.\ B {\bf 261}, 1 (1985).
}
%
\lref\ftb{
E.~S.~Fradkin and A.~A.~Tseytlin,
``Nonlinear Electrodynamics From Quantized Strings,''
Phys.\ Lett.\ B {\bf 163}, 123 (1985).
}
%
\lref\leigh{
R.~G.~Leigh,
``Dirac-Born-Infeld Action From Dirichlet Sigma Model,''
Mod.\ Phys.\ Lett.\ A {\bf 4}, 2767 (1989).
}
%
%
\lref\pol{
J.~Polchinski,
``Dirichlet-Branes and Ramond-Ramond Charges,''
Phys.\ Rev.\ Lett.\  {\bf 75}, 4724 (1995)
[arXiv:hep-th/9510017].
}
%
\lref\pop{
M.~Aganagic, C.~Popescu and J.~H.~Schwarz,
``Gauge-invariant and gauge-fixed D-brane actions,''
Nucl.\ Phys.\ B {\bf 495}, 99 (1997)
[arXiv:hep-th/9612080].
}
%
\lref\ceda{
M.~Cederwall, A.~von Gussich, B.E.W.~Nilsson and A.~Westerberg,
``The Dirichlet super-three-brane in ten-dimensional type IIB  supergravity,''
Nucl.\ Phys.\ B {\bf 490}, 163 (1997)
[arXiv:hep-th/9610148].
}
\lref\aga{
M.~Aganagic, C.~Popescu and J.~H.~Schwarz,
``D-brane actions with local kappa symmetry,''
Phys.\ Lett.\ B {\bf 393}, 311 (1997)
[arXiv:hep-th/9610249].
}
%
\lref\cedb{
M.~Cederwall, A.~von Gussich, B.E.W.~Nilsson, P.~Sundell and A.~Westerberg,
``The Dirichlet super-p-branes in ten-dimensional type IIA and IIB  supergravity,''
Nucl.\ Phys.\ B {\bf 490}, 179 (1997)
[arXiv:hep-th/9611159].
}
%
\lref\bt{
E.~Bergshoeff and P.~K.~Townsend,
``Super D-branes,''
Nucl.\ Phys.\ B {\bf 490}, 145 (1997)
[arXiv:hep-th/9611173].
}
%
%
\lref\brs{
E.~Bergshoeff, M.~Rakowski and E.~Sezgin,
``Higher Derivative Superyang-Mills Theories,''
Phys.\ Lett.\ B {\bf 185}, 371 (1987).
}
%
%
\lref\gv{
S.~J.~Gates and S.~Vashakidze,
``On D = 10, N=1 Supersymmetry, Superspace Geometry And Superstring Effects,''
Nucl.\ Phys.\ B {\bf 291}, 172 (1987).
}
%
\lref\sven{
S.~F.~Kerstan,
``Supersymmetric Born-Infeld from the D9-brane,''
arXiv:hep-th/0204225.
}
%
%
\lref\fks{
L.~De Fosse, P.~Koerber and A.~Sevrin,
``The uniqueness of the Abelian Born-Infeld action,''
Nucl.\ Phys.\ B {\bf 603}, 413 (2001)
[arXiv:hep-th/0103015].
}
%
\lref\hrs{
P.~S.~Howe, O.~Raetzel and E.~Sezgin,
``On brane actions and superembeddings,''
JHEP {\bf 9808}, 011 (1998)
[arXiv:hep-th/9804051].
}
%
%
\lref\nic{
N.~Wyllard,
``Derivative corrections to the D-brane Born-Infeld action: Non-geodesic  embeddings and the Seiberg-Witten map,''
JHEP {\bf 0108}, 027 (2001)
[arXiv:hep-th/0107185].
}
%
\lref\cor{
L.~Cornalba,
``Corrections to the Abelian Born-Infeld 
action arising from  noncommutative geometry,''
JHEP {\bf 0009}, 017 (2000)
[arXiv:hep-th/9912293].
}
%
\lref\das{
S.~R.~Das, S.~Mukhi and N.~V.~Suryanarayana,
``Derivative corrections from noncommutativity,''
JHEP {\bf 0108}, 039 (2001)
[arXiv:hep-th/0106024].
}
%
\lref\sw{
N.~Seiberg and E.~Witten,
``String theory and noncommutative geometry,''
JHEP {\bf 9909}, 032 (1999)
[arXiv:hep-th/9908142].
}
%
\lref\tsa{
A.~A.~Tseytlin,
``Renormalization Of 
Mobius Infinities And Partition 
Function Representation For String Theory Effective Action,''
Phys.\ Lett.\ B {\bf 202}, 81 (1988).
}
%
\lref\loops{
A.~A.~Tseytlin,
``Vector Field Effective Action In The Open Superstring Theory,''
Nucl.\ Phys.\ B {\bf 276}, 391 (1986)
[Erratum-ibid.\ B {\bf 291}, 876 (1987)].
}
%
\lref\bil{
A.~Bilal,
``Higher-derivative corrections to the non-abelian Born-Infeld action,''
Nucl.\ Phys.\ B {\bf 618}, 21 (2001)
[arXiv:hep-th/0106062].
}
%
%
\lref\kita{
Y.~Kitazawa,
``Effective Lagrangian For Open Superstring From Five Point Function,''
Nucl.\ Phys.\ B {\bf 289}, 599 (1987).
}
%
\lref\refolli{
A.~Refolli, A.~Santambrogio, N.~Terzi and D.~Zanon,
``$F^5$ contributions to the nonabelian Born Infeld action from a  
supersymmetric Yang-Mills five-point function,''
Nucl.\ Phys.\ B {\bf 613}, 64 (2001)
[arXiv:hep-th/0105277].
}
%
%
\lref\sevr{
P.~Koerber and A.~Sevrin,
``The non-Abelian Born-Infeld action through order $\alpha^{\prime 3}$,''
JHEP {\bf 0110}, 003 (2001)
[arXiv:hep-th/0108169].
}
\lref\groningen{
A.~Collinucci, M.~ de Roo, M.G.C.~Eenink, 
``Supersymmetric Yang-Mills theory at order $\alpha^{\prime 3}$,''
[arXiv:hep-th/0205150].
}
%
%
\lref\sugra{
M.~Cederwall, U.~Gran, M.~Nielsen and B.E.W.~Nilsson,
``Manifestly supersymmetric M-theory,''
JHEP {\bf 0010}, 041 (2000)
[arXiv:hep-th/0007035]; 
``Generalised 11-dimensional supergravity,''
arXiv:hep-th/0010042.
}
\lref\berk{
M.~Berkovits, V.~Pershin,
``Supersymmetric Born-Infeld from the Pure Spinor 
Formalism of the Open Superstring,''
[arXiv:hep-th/0205154] }


%
\Title{\vbox{\baselineskip12pt
\hbox{G{\"o}teborg ITP preprint}
\hbox{hep-th/0205165} }} 
{\vbox{ \centerline{Spinorial cohomology of} 
\centerline {abelian $d=10$ super-Yang-Mills at $\co(\alpha^{\prime 3})$}
}}

\bigskip
\centerline{Martin Cederwall, 
Bengt E.W.~Nilsson and Dimitrios
Tsimpis}

\bigskip

\centerline{{\sl Department of Theoretical Physics}}
\centerline{\sl G{\"o}teborg University and Chalmers University of
Technology}
\centerline{\sl SE-412 96 G{\"o}teborg, Sweden}
\centerline{\tt tfemc, tfebn, tsimpis@fy.chalmers.se}

\bigskip
\centerline{\bf Abstract}
\medskip

\noindent We compute the spinorial cohomology 
of ten-dimensional abelian SYM at order $\alpha^{\prime 3}$ 
and we find that it is trivial. Consequently,
linear supersymmetry {\it alone} excludes the presence
of $\alpha^{\prime 3}$-order corrections. 
Our result lends support to the
conjecture that there may be a unique
supersymmetric deformation of ordinary ten-dimensional
abelian SYM.

\Date{May 2002}

\vfill\eject

\newsec{Introduction}

It was realized some time ago that 
the Born-Infeld action \bi\ 
(see \tseytlin\ for a review)
occurs in
string theory as the 
tree-level open-string effective action
in the limit
of slowly varying field-strengths \refs{\fta, \ftb}.
After the discovery of D-branes \pol,
the BI action
in $p+1$ spacetime
dimensions was re-interpreted
as the low-energy effective theory
on the world-volume of a single $p$-brane \leigh.
The supersymmetrization of the BI in ten dimensions
was obtained in \pop\ by gauge-fixing the
$\kappa$-symmetric D-brane actions of 
\refs{\ceda, \aga, \cedb, \bt}. It has been
subsequently re-derived
using the superembedding
formalism \hrs\ \foot{See \berk\ for
a recent discussion in the context of 
the pure-spinor formalism of open superstrings.}. 

There have been claims in the literature 
that due to the restrictive form of 
supersymmetry in ten dimensions,  the abelian 
supersymmetric BI 
may be the unique supersymmetric deformation of ordinary 
abelian SYM, in the limit of slowly-varying field-strengths.
There is evidence in favor of this claim coming from
the Seiberg-Witten map \sw\ and the 
``form-invariance'' of the BI action under it \refs{\cor, \nic},
as well as from studying the deformations
of certain BPS configurations of D-branes \fks.
A stronger form of this conjecture would be that there is a unique
supersymmetric deformation of ordinary abelian SYM, even 
when the slowly-varying field-strength assumption $\pa F =0$ is lifted.
The abelian BI would then be the $\pa F\rightarrow 0$ limit
of that unique deformation.

In this paper we lend support to this 
conjecture: By using superspace techniques, we investigate
the space of all possible supersymmetric deformations of 
ten-dimensional abelian SYM at order $\alpha^{\prime 3}$ and
we find that it is zero-dimensional. Consequently, to this order
in $\alpha^{\prime}$, the ten-dimensional abelian BI is indeed the unique 
supersymmetric deformation of ordinary abelian SYM.

A related result was proven some time ago \at\  
(see also \tsa) 
in the context of open-string theory.
One of the implications 
of this reference is 
that at tree level in the string coupling constant 
there are no derivative corrections to the 
ten-dimensional abelian 
supersymmetric BI at order $\a^{\prime 3}$, in contrast to
the bosonic case. Our result makes no reference to 
string theory. The only input is ten-dimensional {\it linear} 
supersymmetry.

The superspace Bianchi identities for ten-dimensional SYM were solved 
in \cnta, 
using only the standard 
conventional constraint \refs{\ga, \gb}. 
It was observed that the deformations 
of ordinary SYM are controlled (parameterized) by a chiral five-form
$J_{5}$  (see 
\refs{\gv, \brs} for earlier work). 
This object may be thought of as a composite operator
made of the SYM fields $\l^\a$ (the gaugino) and $F_{ab}$ 
(the gauge-invariant field-strength). One can obtain 
an $\a^{\prime}$ expansion, by
considering  the most general ansatz for $J_{5}$
at each order in $\a^{\prime}$. By inserting the ansatz 
into the Bianchi identities, the tower of $\a^{\prime}$
corrections is generated. 
This procedure was carried out to order $\a^{\prime 2}$ 
in \cntb.

However, as noted in \cnta, not every $J_{5}$ is consistent with the 
Bianchi identities. The consistency condition $D_2J_{5}=0$ 
has to be satisfied, where the action of $D_2$ is given by 
a spinor superderivative and 
a subsequent projection onto the highest representation. Moreover,
some part of $J_{5}$ will generically be removable 
by redefinitions of the spinor superfield $A_\a$. These redefinitions
shift $J_{5}$ by an amount $D_1A_\a$, where the action 
of $D_1$ is again given by a spinor superderivative and
a projection onto the highest representation. 
The effective deformations of the theory are therefore
given by the most general $J_5$ which 
cannot be written as $D_1A_\a$ and obeys $D_2J_5=0$.
As remarked in  \refs{\martin, \cntc} the operators
$D_{1,2}$ satisfy a nilpotency condition
\eqn\nilpot{D_2\circ D_1=0}
and define a spinorial cohomology $SH$,
\eqn\scoh{SH=Ker(D_2)/Im(D_1).}
The supersymmetric deformations of the theory are 
in one-to-one correspondence with elements of $SH$.
 
We should emphasize that the {\it only} assumption in this procedure 
is that the theory possesses linear
supersymmetry. This is to be contrasted with
the superembedding formalism or $\k$-symmetry, which ``know''
about nonlinear supersymmetry as well and are therefore more restrictive 
in principle.

In this paper we compute $SH$ for $N=1$, $d=10$ abelian SYM at order 
$\a^{\prime 3}$ , counting orders of
$\a^{\prime}$ relative to the $F^2$ term. 
We find that $SH=0$ and 
in view of the remark
following \scoh, we conclude that there is no possible supersymmetric 
deformation of ordinary abelian SYM at this order in $\alpha^{\prime}$.

The next section is a brief review of SYM
in the language of ten-dimensional superspace. Section 3 contains
the computation of spinorial cohomology
at order  $\alpha^{\prime 3}$. We conclude with some discussion 
in section 4.

\newsec{Review of $d=10$ SYM.}

This section contains a brief review of ten-dimensional SYM in 
superspace language, mainly for establishing notation
and conventions. For a more extensive discussion the reader
is referred to \refs{\cnta, \cntb}.

Ten-dimensional SYM contains a gauge superfield $(A_\a, A_a)$. 
Ordinary SYM is obtained by imposing the constraint \nilssona,
\eqn\ordinarysym{F_{\a\b}=0.}
The vector part of this constraint, $\G_a^{\a\b}F_{\a\b}=0$,
is the so called  conventional constraint
\refs{\ga, \gb} and  serves the purpose of eliminating a redundant vector
potential sitting at first level in the $\theta$ expansion of $A_\a$.
In order to relax \ordinarysym\ one expresses 
$F_{\a\b}$ in terms of an anti-self-dual five-form $J_5$,
\eqn\relaxedsym{F_{\a\b}={1\o 5!}\G^{a_1\dots a_5}_{\a\b}J_{a_1\dots a_5}.}
Note that the conventional constraint is still implemented.

When \relaxedsym\ is plugged into the superspace Bianchi identities,
the (relaxed) equations of motion for the gaugino $\l^\a$ and
the gauge-invariant field strength $F_{ab}$ are obtained in terms
of fields appearing at various levels in the $\theta$ expansion of
$J_5$. The superspace Bianchi identities 
with the relaxed constraint \relaxedsym\ 
were solved in \cnta. In the limit
$J_5\rightarrow 0$ one recovers ordinary SYM. 
By expanding $J_5$ in $\alpha^\prime$, one
generates the tower of $\alpha^\prime$ corrections to
ordinary SYM. As mentioned in the introduction,
the Bianchi identities imply that $J_5$ satisfies,
\eqn\diduo{D_2 J_{a_1\dots a_5}:=D_\a J_{a_1\dots a_5}\vert =0,}
where by $\vert$ we denote the projection onto the 
anti-self-dual $(00030)$ part. An explicit expression 
for the projection is given
below in equation (3.9). Moreover, a redefinition $\d A_\a$
of the spinor superpotential shifts $J_5$ by an 
amount proportional to $D_1\d A_\a$, where,
\eqn\dieva{D_1 \d A_\a :={1\o 5!}\G_{a_1\dots a_5}^{\a\b}
D_{\a}\d A_{\b} . }
Supersymmetric deformations of the theory are elements of 
the spinorial cohomology \scoh.

Let us conclude this section with some comments 
on dimensions and representations. 
The mass dimensions of 
the various fields are as follows:
\eqn\engdims{[A_{\a}]={1\o 2};
\,\,\,\,\,[\l^\a]={3\o 2}; \,\,\,\,\,[F_{ab}]=2; 
\,\,\,\,\,[J_{a_1\dots a_5}]=1, }
while $[\a^\prime]=-2$.
In terms of representations of $so(1,9)\approx  D_5$ 
the fields transform as
\eqn\reps{A_{\a}\sim (00010);
\,\,\,\,\, \l^\a\sim (00001); \,\,\,\,\, F_{ab}\sim (01000); 
\,\,\,\,\, J_{a_1\dots a_5}\sim (00020). }
Our convention is that $(00020)$ is anti-self-dual whereas 
$(00002)$ is self-dual. Using the lowest-order  equations of motion
\eqn\loeqs{\G^aD_a\l=0;\,\,\,\,\, D^aF_{ab}=0,}
it is not difficult to see that 
\eqn\morereps{D_{a_1}\dots D_{a_n}\l\sim (n0001);\,\,\,\,\,
D_{a_1}\dots D_{a_n}F_{ab}\sim (n1000).  }
Finally, the action of the spinor superderivative to
lowest order in $\alpha^\prime$ is given by,
\eqn\supder{D_\a \l^\b={1\o 2}(\G^{ab})_\a{}^\b F_{ab};\,\,\,\,\,
D_\a F_{ab}=2(\G_{[a} D_{b]}\l)_\a. }

\newsec{Spinorial cohomology at ${\cal O}(\alpha^{\prime 3})$}

In order to compute the spinorial cohomology at 
${\cal O}(\alpha^{\prime 3})$, we have to carry out the
following  procedure: a) Write down the most general
anti-self-dual five-form $J_5$ 
and spinor-potential redefinition $\d A_\a$
which are allowed by
dimensional analysis
and are of the form $\alpha^{\prime 3}$ times a product of
$\l^\a$'s and $F_{ab}$'s. b) Compute the action of $D_{1,2}$. 
c) Compute the spinorial cohomology using \scoh.
Let us note that in 
carrying out steps a) -- c) above, one need only use
the {\it lowest-order} equations \loeqs, \supder.
This is because $\d A_\a, \,\, J_5$ already contain one power
of $\alpha^{\prime 3}$.

Taking the last paragraph of the previous section into account, 
we find that the most general $J_{5}$ at order 
$\a^{\prime 3}$ is of the form
\eqn\generaljfive{J_{a_1\dots a_5}
=\sum_{i=1}^{10}b_i B^{(i)}_{a_1\dots a_5}, }
where $b_1,\dots b_{10} \in \IR$.
Schematically,  
$B^{(1)}\dots B^{(10)}$ are given by the
projections of the following 
products of irreducible 
representations onto $(00020)$:
\eqn\bidef{\eqalign{
B^{(1)}&\sim F^2\l^2 \sim (01000)^2_s\otimes(00001)^2_a\cr
B^{(2)}&\sim \l^3D\l\sim  (00001)^3_a\otimes (10001)\cr
B^{(3)}&\sim F^2DF\sim (01000)^2_s\otimes(11000)\cr
B^{(4,5,6)}&\sim F(D\l)^2\sim (01000)\otimes (10001)^2\cr
B^{(7)}&\sim F\l D^2\l\sim (01000)\otimes (00001)\otimes (20001)\cr
B^{(8,9,10)}&\sim DF\l D\l\sim (11000)\otimes (00001)\otimes (10001).
\cr}}
More explicitly (antisymmetrization in $a_1\dots a_5$ is implied
on the right-hand side),
\eqn\expbidef{\eqalign{
B^{(1)}_{a_1\dots a_5}&=(\l\G_{a_2\dots a_5}{}^{ijk}\l) F_{a_1i}F_{jk}
-{\rm dual}\cr
B^{(2)}_{a_1\dots a_5}&=2(\l\G_{a_1}D_{a_2}\l)
(\l\G_{a_3a_4a_5}\l)-{\rm dual}\cr
B^{(3)}_{a_1\dots a_5}&=F_{a_1a_2}F_{a_3i}D^iF_{a_4a_5}
-{\rm dual}\cr
B^{(4)}_{a_1\dots a_5}&=(D_{a_1}\l\G_{a_2a_3a_4}D^i\l)F_{a_5i}-{\rm dual}\cr
B^{(5)}_{a_1\dots a_5}&=(D^i\l\G_{a_1a_2a_3}D_i\l)F_{a_4a_5}
-{\rm dual}\cr
B^{(6)}_{a_1\dots a_5}&=(D_{a_1}\l\G_{a_2}D_{a_3}\l)F_{a_4a_5}-{\rm dual}\cr
B^{(7)}_{a_1\dots a_5}&=(D^iD_{a_1}\l\G_{a_2a_3a_4}\l)F_{a_5i}
-{\rm dual}\cr
B^{(8)}_{a_1\dots a_5}&=(\l\G_{a_1a_2i}D_{a_3}\l)D^iF_{a_4a_5}-{\rm dual}\cr
B^{(9)}_{a_1\dots a_5}&=(\l\G_{a_1a_2a_3a_4i}D_{j}\l)
D^iF^j{}_{a_5}-{\rm dual}\cr
B^{(10)}_{a_1\dots a_5}&=(\l\G_{a_1a_2a_3}D^{i}\l)
D_iF_{a_4a_5}-{\rm dual}.
\cr}}
A factor of $\a^{\prime 3}$ is suppressed
on the right-hand side above and in the following.

A similar analysis can be performed for the 
most general redefinition $\d A_\a$ at order $\a^{\prime 3}$.
The result is,
\eqn\generalredef{\d A_{\a}
=\sum_{i=1}^{8}a_i A^{(i)}_{\a}, }
where $a_1,\dots a_{8} \in \IR$. Schematically,  
$A^{(1)}\dots A^{(8)}$ are given by the
projections of the following 
products of irreducible 
representations onto $(00010)$:
\eqn\adef{\eqalign{
A^{(1)}&\sim F\l^3 \sim (01000)\otimes(00001)^3_a\cr
A^{(2,3)}&\sim \l(D\l)^2\sim  (00001)\otimes (10001)^2\cr
A^{(4,5)}&\sim FDF\l\sim (01000)\otimes(11000)\otimes(00001)\cr
A^{(6,7)}&\sim F^2(D\l)\sim (01000)^2_s\otimes (10001)\cr
A^{(8)}&\sim DF D^2\l\sim (11000)\otimes (20001).
\cr}}
Explicitly,
\eqn\expadef{\eqalign{
A^{(1)}_{\a}&=F^{ij}(\G^{k}\l)_\a (\l\G_{ijk}\l)\cr
A^{(2)}_{\a}&=(\G^{ijk}\l)_\a (D_i\l\G_{j}D_k\l)\cr
A^{(3)}_{\a}&=(\G^{ijk}\l)_\a (D^l\l\G_{ijk}D_l\l)\cr
A^{(4)}_{\a}&=(\G^{k}\l)_\a D_kF_{ij}F^{ij} 
+ (\G^{ijk}\l)_\a D^lF_{ij}F_{lk}\cr
A^{(5)}_{\a}&=-{7\o 4}(\G^{k}\l)_\a D_kF_{ij}F^{ij} 
+ {1\o 4}(\G^{ijk}\l)_\a D^lF_{ij}F_{lk}\cr
A^{(6)}_{\a}&=(\G^{ijk}D^l\l)_\a F_{ij}F_{kl}\cr
A^{(7)}_{\a}&=(\G^{i}D^j\l)_\a F_{i}{}^lF_{jl}\cr
A^{(8)}_{\a}&=D^iF^{jk}(\G_kD_iD_j\l)_\a.
\cr}}
Finally, the 
most general 
anti-self-dual five-form spinor $\tilde{J}_{a_1\dots a_5\a}$ 
at order $\a^{\prime 3}$ is of the form,
\eqn\genconst{\tilde{J}_{a_1\dots a_5\a}
=\sum_{i=1}^{4}c_i C^{(i)}_{a_1\dots a_5\a,}}
where $c_1,\dots c_{4} \in \IR$. Schematically,  
$C^{(1)}\dots C^{(4)}$ are given by the
projections of the following 
products of irreducible 
representations onto $(00030)$:
\eqn\adef{\eqalign{
C^{(1)}&\sim F\l^2D\l \sim (01000)\otimes(00001)^2_a\otimes(10001)\cr
C^{(2)}&\sim FDFD\l\sim  (01000)\otimes(11000)\otimes (10001)\cr
C^{(3)}&\sim \l D\l D^2\l\sim (00001)\otimes(10001)\otimes(20001)\cr
C^{(4)}&\sim (D\l)^3\sim (10001)^3.
\cr}}
Note that in the decomposition of the above products into irreducible 
representations, there is no $(00003)$ component. Therefore the projection
$\tilde{S}_{a_1\dots a_5\a}\vert$ 
of a spinor-five-form $\tilde{S}_{a_1\dots a_5\a}$ 
onto its gamma-traceless part is equivalent to
the projection onto the $(00030)$ (the anti-self-dual) part.
We have,
\eqn\theprojection{\eqalign{
\tilde{S}_{a_1\dots a_5}\vert&=
{1\o 6}\tilde{S}_{a_1\dots a_5}
-{1\o 6}\G_{a_1}{}^i\tilde{S}_{a_2\dots a_5i}\cr
&-{1\o 12}\G_{a_1a_2}{}^{ij}\tilde{S}_{a_3a_4a_5ij}
+{1\o 36}\G_{a_1a_2a_3}{}^{ijk}\tilde{S}_{a_4a_5ijk}\cr
&+{1\o 144}\G_{a_1\dots a_4}{}^{ijkl}\tilde{S}_{a_5ijkl}
-{1\o 720}\G_{a_1\dots a_5}{}^{ijklm}\tilde{S}_{ijklm},
\cr}
}
where antisymmetrization in 
$a_1\dots a_5$ is understood 
on the right-hand-side and we have omitted all spinor indices.
We can now give an explicit form for $C^{(1)}\dots C^{(4)}$,
\eqn\expadef{\eqalign{
C^{(1)}_{\a}&=(\G^{i}D_{a_1}\l)_\a(\l\G_{a_2a_3a_4}\l)F_{a_5i}\vert \cr
C^{(2)}_{\a}&=(\G^{i}D_{a_1}\l)_\a D_iF_{a_2a_3}F_{a_4a_5}\vert\cr
C^{(3)}_{\a}&=(\G^{i}D_{a_1}\l)_\a(\l\G_{a_2a_3a_4}D_{a_5}D_i\l)\vert\cr
C^{(4)}_{\a}&=(\G^{i}D_{a_1}\l)_\a(D_{a_2}\l\G_{a_3a_4a_5}D_i\l)\vert.
\cr}}
The action of the operator $D_2$ can be computed in a straight-forward 
way using definition \diduo\ and equations \loeqs, \supder. 
We find,
\eqn\dduoim{\eqalign{
D_2B^{(1)}&=16C^{(1)}\cr
D_2B^{(2)}&=-8C^{(1)}\cr
D_2B^{(3)}&=-2C^{(2)}\cr
D_2B^{(4)}&=2C^{(4)}\cr
D_2B^{(5)}&=0\cr
D_2B^{(6)}&=4C^{(2)}\cr
D_2B^{(7)}&=-2C^{(3)}\cr
D_2B^{(8)}&=-8C^{(2)}\cr
D_2B^{(9)}&=-16C^{(3)}\cr
D_2B^{(10)}&=0.\cr}}
We see that, 
with respect to  the basis $(B^{(1)},\dots B^{(10)})$, 
the Kernel of the operator
$D_2$ is the six-dimensional subspace given by,
\eqn\kernel{\eqalign{Ker(D_2)=
Span\{(b_1,2b_1, 2b_6-4b_8,0,b_5,&b_6,-8b_9,b_8,b_9,b_{10})\}, \cr
&b_1,b_5,b_6,b_8,b_9,b_{10}\in\IR.\cr}}
Similarly, the action of the operator $D_1$ can be computed 
taking \dieva\ into account. 
The result is,
\eqn\dduoim{\eqalign{
D_1A^{(1)}&=-{1\o 2}B^{(1)}-B^{(2)}\cr
D_1A^{(2)}&=-8B^{(6)}-4B^{(8)}+{2\o 3}B^{(10)} \cr
D_1A^{(3)}&=-8B^{(5)}+8B^{(10)}\cr
D_1A^{(4)}&=8B^{(3)}-2B^{(8)}\cr
D_1A^{(5)}&=2B^{(3)}
+{4\o 3}B^{(7)}-{1\o 2}B^{(8)}-{1\o 6}B^{(9)}+{1\o 3}B^{(10)}\cr
D_1A^{(6)}&=-8B^{(3)}-4B^{(6)}\cr
D_1A^{(7)}&=-{1\o 6}B^{(5)}\cr
D_1A^{(8)}&=0.\cr}}
With respect to  the basis $(B^{(1)},\dots B^{(10)})$, 
the Image of the operator
$D_1$ is the six-dimensional subspace given by,
\eqn\image{\eqalign{Im(D_1)=
Span\{(-{1\o 2}a_1,-a_1, 8a_4+2a_5-8a_6,0,-8a_3-{1\o 6}a_7&,\cr
-8a_2-4a_6,{4\o 3}a_5,-4a_2-2a_4-{1\o 2}a_5,-{1\o 6}a_5&,
{2\o 3}a_2+8a_3+{1\o 3}a_5  )\}, \cr
&a_1\dots a_8\in\IR.\cr}}
Comparing \image, \kernel, 
it is easy to see that the Image of $D_1$ is in the 
Kernel of $D_2$. This of course is just the
nilpotency property \nilpot. 

The spinorial 
cohomology $SH$ can be readily computed from \image, \kernel,
\eqn\spincoh{SH=Ker(D_2)/Im(D_1)=0.}
The triviality of $SH$ means that at order 
$\a^{\prime 3}$, there {\it cannot} be
any supersymmetric deformation of ordinary abelian $d=10$ SYM.

%
%
%
%
\bigskip
\epsfxsize=.95\hsize
\epsffile{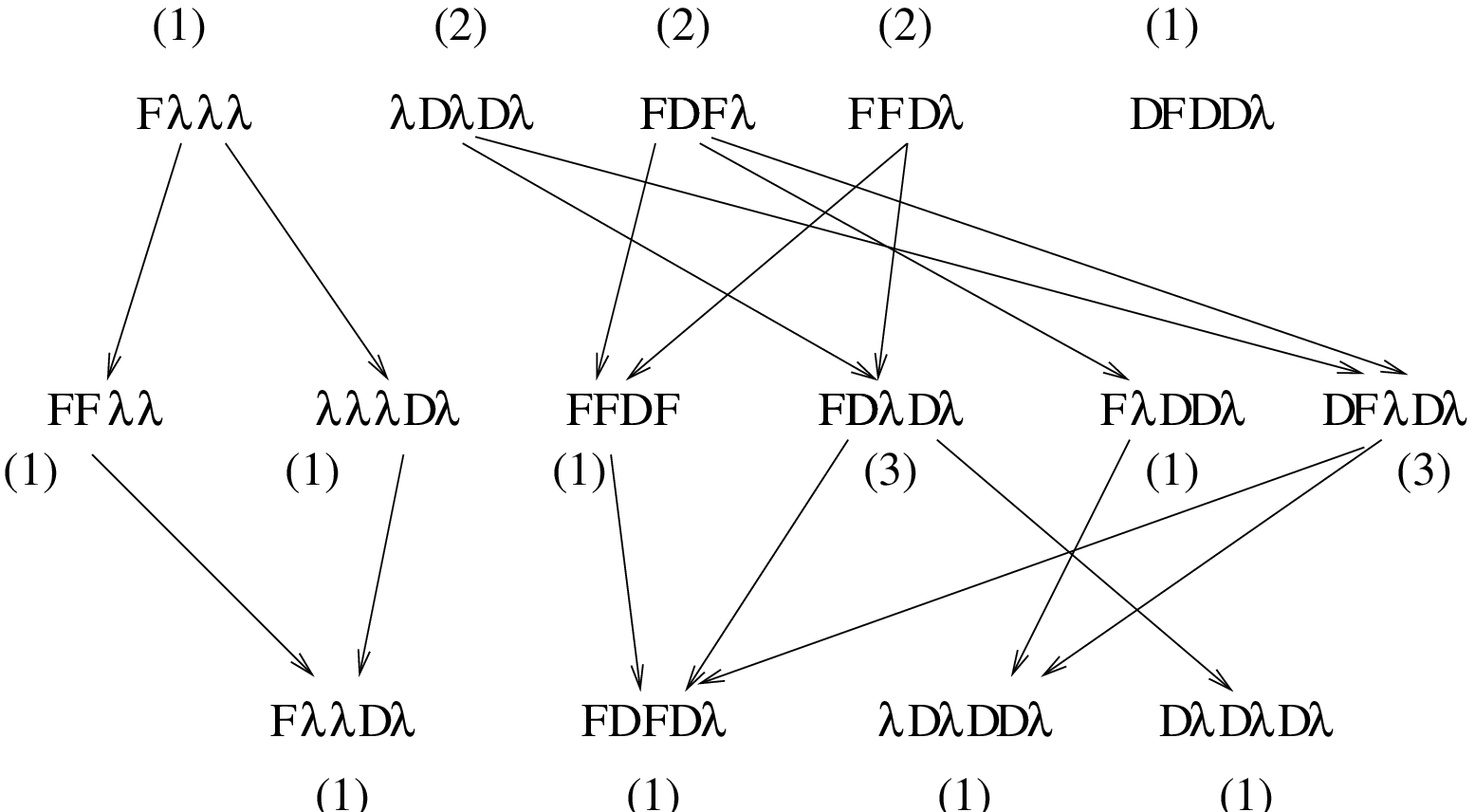}
{\it \noindent Figure 1. In the three rows 
we have included, schematically, 
the various terms in $\d A_\a$, $J_5$, $\tilde{J}_5$
(equations \generalredef, \generaljfive, \genconst,
respectively). The arrows from the first
to the second row indicate the action of $D_1$,
while the arrows from the second to the third row
indicate the action of $D_2$. The numbers in parentheses 
denote multiplicities.}
\bigskip

\newsec{Discussion}

In this paper we computed the spinorial cohomology
for ten-dimensional abelian SYM at order $\a^{\prime 3}$ and
we found that it is trivial. We have therefore demonstrated
that to this order in $\a^{\prime}$ the abelian BI
is the unique supersymmetric deformation of
ordinary abelian SYM.

We would like to emphasize that the only assumption
we make is that the theory 
possesses {\it linear} ten-dimensional supersymmetry.
Consequently, 
any computation coming from superstrings would necessarily
respect \spincoh\ {\it to all orders in the string coupling}.
In particular the 
result of the present paper
 combined with the result of \cntb\  
 implies that, up to order $\a^{\prime 3}$, 
all $g_s$ corrections 
to the abelian supersymmetric BI action 
can be absorbed in an 
overall string-tension renormalization. To
one-loop order in  $g_s$, this was
indeed shown to be the case in \loops.

A natural generalization of this paper 
would be to compute the non-abelian spinorial
cohomology at $\alpha^{\prime 3}$--order 
\foot{See references \refs{\bil, \refolli, \sevr} 
for some (partial) results on the non-abelian BI
at this order in $\alpha^{\prime}$. The bosonic 
part of the ${\cal O} (\alpha^{\prime 3})$
corrections to the effective action of 
type I superstring theory at tree-level in the string coupling
was presented in \kita. The action 
including fermions up to quadratic order
was recently given in \groningen.}. 
This task is very much involved technically. One can 
appreciate the level of complication by noting that 
in the non-abelian case one has to deal
with products of fields valued in the adjoint
of a general Lie group $G$. These products are 
contracted with $G$-invariant tensors of rank as high as five.
The analysis of this paper would have to 
be repeated separately for each plethysm of each 
$G$-invariant tensor and there are generically 
twenty-six non-vanishing plethysms in each
fifth-rank tensor! 

Another 
more feasible, perhaps, generalization 
is the computation of the abelian spinorial cohomology
at order $\alpha^{\prime 4}$.  
It would be interesting to see whether the vanishing of $SH$
persists to this order. Note that $SH=0$ 
would imply that there is no new 
independent superinvariant
arising at order $\alpha^{\prime 4}$; it would
{\it not} imply
$F_{\a\b}=0$ (Equivalently: $J_5=0$). 
In \refs{\cnta, \cntb}, the most general 
$F_{\a\b}$ allowed by linear supersymmetry
was determined to order $\alpha^{\prime 2}$.
As was explained in these references,
the constraint 
$D_2J_5=0$ 
is satisfied to order $\alpha^{\prime 2}$, but 
not to order $\alpha^{\prime 4}$.
To ensure that the constraint is satisfied
to this order, one would need to compensate by 
adding an $\alpha^{\prime 4}$ correction 
to $F_{\a\b}$. 
The $\pa F\rightarrow 0$ limit of this correction
should coincide (presumably after field redefinitions) 
with the recent result of \sven.

The techniques of this paper can also be applied to 
eleven-dimensional supergravity \sugra. 
It would be interesting to obtain
the spinorial cohomology for the first few orders in 
the Planck-length (long-wavelength) expansion and
thereby determine the possible deformations of the theory.
\bigskip
\centerline{\bf Acknowledgments}\nobreak
\bigskip

\noindent 
We would like to thank P.~Howe and N.~Wyllard
for email correspondence. 
This work is supported in part by EU contract
HPRN-CT-2000-00122 and by the Swedish 
Research Council. The programs LiE \lie\
and GAMMA \ulf\ have been very useful for representation-theoretical
considerations and Gamma-matrix manipulations, respectively.

\vfil\eject

\listrefs
\end